%% file: main.tex
\PassOptionsToPackage{table,xcdraw,dvipsnames}{xcolor}

\documentclass[conference]{IEEEtran}
\IEEEoverridecommandlockouts
\usepackage{cite}
\usepackage{amsmath,amssymb,amsfonts}
\usepackage{booktabs} 

\usepackage{algorithmic}
\usepackage{graphicx}
\usepackage{textcomp}
\usepackage[a4paper, total={184mm,239mm}]{geometry}
\usepackage{array}
\usepackage{tabu}

\usepackage{multirow}
\usepackage{etoolbox}
\usepackage{wasysym}
\usepackage{makecell}
\usepackage{tikz}
\usepackage{fontawesome}
\usepackage{enumitem}
\usepackage{tabularx}
\usepackage{float}

\usepackage{xcolor}

\definecolor{gold}{RGB}{255,215,0}    
\definecolor{silver}{RGB}{192,192,192} 
\definecolor{bronze}{RGB}{205,127,50}  

\usepackage{pifont}
\newcommand{\cmark}{\ding{51}}%
\newcommand{\xmark}{\ding{55}}%

\usepackage{listings}
\usepackage{caption}
\DeclareCaptionFont{white}{\color{white}}
\DeclareCaptionFormat{listing}{\colorbox{gray}{\parbox{\textwidth}{#1#2#3}}}
\usepackage{verbatim} 
\usepackage{fancyvrb}
\usepackage{acronym}
\usepackage{amsthm}
\VerbatimFootnotes 

\newcommand{\rot}[1]{%
  \text{\rotatebox[origin=c]{90}{#1}}%
}

\usepackage{tikz}
\newcommand{\ballnumber}[1]{\tikz[baseline=(myanchor.base)] \node[circle,fill=.,inner sep=1pt] (myanchor) {\color{-.}\bfseries\footnotesize #1};}

\usepackage{xspace}

\renewcommand{\paragraph}[1]{\vspace{.4em}\noindent\textbf{#1}}

\makeatletter
\DeclareRobustCommand\onedot{\futurelet\@let@token\@onedot}
\def\@onedot{\ifx\@let@token.\else.\null\fi\xspace}

\def\eg{\emph{e.g}\onedot} 
\def\ie{\emph{i.e}\onedot} 
 
\def\etc{\emph{etc}\onedot} \def\versus{\emph{vs}\onedot}

\makeatother

\newcommand{\sysname}[0]{{\textsc{HaVen}}\xspace}
\input{math_commands}

\usepackage{hyperref}
\hypersetup{
    colorlinks=true,
    citecolor=blue,
    linkcolor=blue,
    filecolor=cyan,      
    urlcolor=magenta,
    pdfpagemode=FullScreen,
    }

\usepackage[capitalise]{cleveref}

\usepackage{fontawesome}

\makeatletter
\patchcmd{\@makecaption}
  {\scshape}
  {}
  {}
  {}
\makeatother
\def\BibTeX{{\rm B\kern-.05em{\sc i\kern-.025em b}\kern-.08em
    T\kern-.1667em\lower.7ex\hbox{E}\kern-.125emX}}

\DeclareUnicodeCharacter{2212}{-}

\begin{document}

\title{\sysname: Hallucination-Mitigated LLM for Verilog Code Generation Aligned with HDL Engineers
\thanks{
This work is partially supported by National Key R\&D Program of China~(2022YFB4500200), National Natural Science Foundation of China~(No.62102257).
Corresponding author: Zhezhi He.
}
}


\author{
    \IEEEauthorblockN{
        Yiyao Yang\IEEEauthorrefmark{1}\IEEEauthorrefmark{4},
        Fu Teng\IEEEauthorrefmark{2}\IEEEauthorrefmark{4},
        Pengju Liu\IEEEauthorrefmark{1}\IEEEauthorrefmark{4},
        Mengnan Qi\IEEEauthorrefmark{1},
        Chenyang Lv\IEEEauthorrefmark{1},
        Ji Li\IEEEauthorrefmark{3},
        Xuhong Zhang\IEEEauthorrefmark{2}
        Zhezhi He\IEEEauthorrefmark{1}}
    \IEEEauthorblockA{\IEEEauthorrefmark{1}School of Electronic Information and Electrical Engineering, Shanghai Jiao Tong University, Shanghai, China}
    \IEEEauthorblockA{\IEEEauthorrefmark{2}School of Software Technology, Zhejiang University, Ningbo, China; \IEEEauthorrefmark{3}Independent Researcher}
    \IEEEauthorblockA{\IEEEauthorrefmark{4}These authors contributed equally}
    \{yangyiyao, zhezhi.he\}@sjtu.edu.cn
}

\maketitle
\begin{abstract}
Recently, the use of large language models (LLMs) for Verilog code generation has attracted great research interest to enable hardware design automation. 
However, previous works have shown a gap between the ability of LLMs and the practical demands of hardware description language (HDL) engineering. 
This gap includes differences in how engineers phrase questions and hallucinations in the code generated.
To address these challenges, we introduce \sysname, a novel LLM framework designed to mitigate hallucinations and align Verilog code generation with the practices of HDL engineers. 
\sysname tackles hallucination issues by proposing a comprehensive taxonomy and employing a chain-of-thought (CoT) mechanism to translate symbolic modalities (\eg truth tables, state diagrams, \etc) into accurate natural language descriptions. 
Furthermore, \sysname bridges this gap by using a data augmentation strategy. It synthesizes high-quality instruction-code pairs that match real HDL engineering practices.
Our experiments demonstrate that \sysname significantly improves the correctness of Verilog code generation, outperforming state-of-the-art LLM-based Verilog generation methods on VerilogEval and RTLLM benchmark.
\sysname is publicly available at \url{https://github.com/Intelligent-Computing-Research-Group/HaVen}.

\end{abstract}


\section{Introduction}
Large language models (LLMs) have achieved remarkable success in generating code for software programming languages like 
Python\cite{luo2023wizardcoderempoweringcodelarge,muennighoff2024octopackinstructiontuningcode, openai2024gpt4, wei2023magicoder, wizardlm2024, zheng2024opencodeinterpreterintegratingcodegeneration}, benefiting from extensive documentation and rich training datasets \cite{github2021}. 
However, fine-tuning LLMs for hardware description language (HDL) generation, 
particularly Verilog, presents distinct challenges that set it apart from software code 
generation\cite{chen2021evaluatinglargelanguagemodels, isyourcodecorrect,liu2023verilogeval}. These challenges arise from several key factors elaborated as follows.


Hallucination\cite{hallu} in language models refers to the generation of incorrect or misleading information not grounded in reality. 
While prior researches, such as CodeHalu\cite{codehalu} and HalluCode\cite{hallucode}, propose taxonomies for hallucination and analyze this phenomenon in the context of Python code generation.
These studies mostly focus on Python-specific error analysis.
In contrast, HDLs exhibit unique characteristics, \eg, their hardware-centric nature, timing requirements, and concurrency models, making them fundamentally different from software languages like Python\cite{li2023starcoder}. 
As a result, LLMs often lack the necessary domain-specific knowledge, resulting in frequent hallucinations, especially when LLMs attempt to generate Verilog code that aligns with the practices of HDL engineers\footnote{Recently, the popular VerilogEval benchmark, recognizing the same issue, proposed a new version that focuses on human evaluation problems.\cite{pinckney2024revisiting}}. They may produce syntax, functional or logical errors that do not match the expectation of HDL engineers.
Previous works like RTLFixer\cite{rtlfixer} and OriGen\cite{2024origen} have focused on \emph{enhancing LLMs by reducing syntax errors}. 
However, these methods fall short in providing the \emph{diverse, domain-rich Verilog samples needed to capture HDL-specific nuances}, and thus do not effectively mitigate hallucinations.


\begin{table}[t]
\centering
\caption{\textbf{LLM-generated vs. HDL engineer-used formats}}
\label{tab:my-table}
\resizebox{\columnwidth}{!}{%
\begin{tabular}{|c|c|}
\hline
\textbf{Format generated by LLM} & \textbf{Format used by HDL engineers} \\ \hline
\begin{tabular}[c]{@{}l@{}}
``Implement the state machine with a combinational always\\ block, which is used to determine the next state based on\\ the current state and the values of the x port. If the current \\ state is A and the x port is low, then..."
\end{tabular} & \begin{tabular}{|c|c|c|c|}
\hline
\textbf{Current State} & \textbf{Input (x)} & \textbf{Next State} & \textbf{Output (out)} \\ \hline
A                     & 0                  & B                   & 0                     \\ \hline
A                     & 1                  & A                   & 0                     \\ \hline
B                     & 0                  & A                   & 1                     \\ \hline
B                     & 1                  & B                   & 1                     \\ \hline
\end{tabular} \\ \hline
\end{tabular}%
}
\end{table}

\begin{table*}[t]
\centering
\caption{\textbf{Taxonomy of Hallucination in LLM-based Verilog
Code Generation}}
\label{tab:hallucination_taxonomy}
\resizebox{\textwidth}{!}{%
\renewcommand{\arraystretch}{1.05}
\begin{tabular}{|c|c|c|l|c|}
\hline
Type & Sub-types & Example prompt causing hallucination & \begin{tabular}[c]{@{\qquad\qquad\qquad}c@{}}Incorrect code \end{tabular}& Error analysis \\ \hline
 \cellcolor[HTML]{F6CECA} 
 &  
 \begin{tabular}[c]{@{}c@{}}State Diagram \\ Misinterpretation\end{tabular} 
 & \begin{tabular}[c]{@{}c@{}}  Implement this FSM... \\ A{[}out=0{]}--{[}in==0{]}--\textgreater{}B...   \\ A{[}out=0{]}--{[}in==1{]}--\textgreater{}A... \end{tabular} & \begin{tabular}[l]{@{}l@{}}\texttt {case(state)}\\ \texttt{\qquad A: begin}\\ \texttt{\qquad \qquad if(in) next\_state = \textcolor{red}{\textbf{B}};}\\ \texttt{\qquad \qquad else next\_state = \textcolor{red}{\textbf{A}};...}\end{tabular} & ``A'' and ``B'' should be reversed. \\ 
 \cline{2-5}  
 \cellcolor[HTML]{F6CECA} 
 
 & 
 \begin{tabular}[c]{@{}c@{}}Waveform Chart \\ Misinterpretation\end{tabular} & \begin{tabular}[c]{@{}c@{}} Implement the waveforms below...\\\; a: 0 1 0 1 ...\\ \; b: 0 0 1 1 ...\\ out: 0 0 0 1 ...\end{tabular} & \begin{tabular}[l]{@{}l@{}}...\\ \texttt{assign out = \textcolor{red}{\textbf{a + b}};}\\ ...\end{tabular} & ``out'' should be ``a \& b". \\ \cline{2-5} 
 \cellcolor[HTML]{F6CECA} 
\multirow{-8}{*}{\begin{tabular}[c]{@{}c@{}}\textbf{Symbolic}\\ \textbf{Hallucination}\end{tabular}} 
\cellcolor[HTML]{F6CECA} 
& \begin{tabular}[c]{@{}c@{}}Truth Table \\ Misinterpretation\end{tabular} & \begin{tabular}[c]{@{}c@{}} Implement the truth table below...\\ a $\mid$ b $\mid$ out\\ \;1 $\mid$ 1 $\mid$ 1 ...   \\ \;1 $\mid$ 0 $\mid$ 0 ...\end{tabular} & {  \begin{tabular}[l]{@{}l@{}}...\\ \texttt{assign out = \textcolor{red}{ \textbf{a $\mid$ b}};}\\ ...\end{tabular}} & {\begin{tabular}[c]{@{}c@{}}
     ``out'' should be ``a \& b''.\\
     (Due to limited space, truth table is partially omitted)
\end{tabular}} 
\\ 
\Xhline{1px}
 \cellcolor[HTML]{C5E5FB}
 
 & 
 \begin{tabular}[c]{@{}c@{}}Digital Design Convention \\ Misapplication\end{tabular} & \begin{tabular}[c]{@{}c@{}}Implement a digit detector, \\using conventional FSM ...\end{tabular} & {  \begin{tabular}[l]{@{}l@{}} \texttt{ case(state)}\\ \texttt{ \qquad A: \textcolor{red}{\textbf{state}}=a+b; ...}\end{tabular}} & \begin{tabular}[c]{@{}c@{}}``state'' should be ``next\_state''. A conventional\\ FSM should contain ``state transition'',\\``next\_state logic'', ``output logic''.\end{tabular} \\ \cline{2-5} 
 \cellcolor[HTML]{C5E5FB}
 & \begin{tabular}[c]{@{}c@{}}Verilog Syntax \\ Misapplication\end{tabular} & Implement a 4-bit adder ... & {  \begin{tabular}[l]{@{}l@{}}\texttt{\textcolor{red}{\textbf{def}} adder\_4bit()}\\... \end{tabular}} & \begin{tabular}[c]{@{}c@{}}The definition of module is syntactically wrong. \\ ``def'' should be ``module''\end{tabular} \\ \cline{2-5} 
\multirow{-5}{*}{ \begin{tabular}[c]{@{}c@{}}\textbf{Knowledge}\\ \textbf{Hallucination}\end{tabular}} 
 \cellcolor[HTML]{C5E5FB}
& 
\begin{tabular}[c]{@{}c@{}}Misunderstanding \\ Verilog-specific Attributes\end{tabular} & \begin{tabular}[c]{@{}c@{}}Implement this module using \\ asynchronous reset signal\end{tabular} & {  \begin{tabular}[l]{@{}l@{}}\texttt{\textcolor{red}{\textbf{always@(posedge clk)}}}\\ \texttt{\qquad if(!reset)...}\end{tabular}} & \begin{tabular}[c]{@{}c@{}}The reset should be asynchronous.\end{tabular} 
\\ 
\Xhline{1px}

\cellcolor[HTML]{FFEEC1} 
 & Incorrect Logical Expression & \begin{tabular}[c]{@{}c@{}} Create a module, the output signal\\  equals a plus b, then or c.\end{tabular} & {  \begin{tabular}[c]{@{}c@{}}\texttt{assign output = \textcolor{red}{\textbf{(a + c) \& b}};}\\ ...\end{tabular}} & ``output'' signal should be ``(a+b)$\mid$c''. \\ \cline{2-5} 
\cellcolor[HTML]{FFEEC1} 
 & \begin{tabular}[c]{@{}c@{}}Incorrect Handling \\ of Corner Cases\end{tabular} & \begin{tabular}[c]{@{}c@{}} Implement logic of two inputs. \\ Output equals 1 when a and\\ b are both 1, otherwise 0.\end{tabular} & {  \begin{tabular}[l]{@{}l@{}}\texttt{\textcolor{red}{\textbf{case}}(\{a, b\})}\\ \texttt{\qquad 2'b11: out = 1};\\ \texttt{endcase}\end{tabular}} & ``default'' case is ignored. \\ \cline{2-5} 
\multirow{-6}{*}{\begin{tabular}[c]{@{}c@{}}\textbf{Logical}\\ \textbf{Hallucination} \end{tabular}} 
\cellcolor[HTML]{FFEEC1} 
& \begin{tabular}[c]{@{}c@{}}Failure in Adhering to \\ Instructional Logic\end{tabular} & \begin{tabular}[c]{@{}c@{}} Implement the logic below:\\ if a == 0 \&\& b == 0; out = 0;\\ elif a == 1 \&\& b == 0; out = 0...\end{tabular} & {  \begin{tabular}[l]{@{}l@{}}\texttt{if(\textcolor{red}{\textbf{a==0 $\mid$$\mid$ b==0}}) out = 0;}\\ \texttt{else if(a==1 \&\& b==0) out = 0;}\\...\end{tabular}} & \begin{tabular}[c]{@{}c@{}}The first ``if'' expression should be \\ ``a==0 \&\& b==0''.\end{tabular} \\ \hline
\multicolumn{5}{l}{
$\dagger$: \texttt{\textcolor{red}{\textbf{code}}} in red highlights the error raised during the code generation.
}
\end{tabular}%
}
\end{table*}

The practices of HDL engineers refer to standardized formats for describing functionality and problems, along with digital design conventions crucial for well-structured code implementation\cite{Badiger2020, Lin2008}.
The scarcity of high-quality datasets aligned with these practices is the primary reason for hallucinations and the lack of domain-specific knowledge in LLMs.
Fine-tuning LLMs for Verilog code generation requires datasets consisting of rich instruction-code pairs, which are currently scarce. 
Although open-source repositories\cite{github_repo} contain Verilog code snippets, these are often devoid of comprehensive natural language descriptions. Existing solutions, including RTLCoder\cite{liu2024rtlcoder}, OriGen\cite{2024origen}, and AutoVCoder\cite{autocoder2024}, have attempted to generate data by using closed-source LLMs to add descriptions to open-source Verilog code samples. However, this approach has notable limitations:
\begin{itemize}[leftmargin=*] 
    \item \textbf{Trivial and Misaligned Descriptions:} Closed-source LLMs are not specialized in HDL. The descriptions generated often fail to capture the complexities of real-world register-transfer level (RTL) design scenarios, as shown in \cref{tab:my-table}.
    
    \item \textbf{Hallucinations in Fine-Tuned Models:}
    Fine-tuning LLMs on such misaligned data can result in models that are prone to hallucination in practical RTL design scenarios. LLMs fail to realize the practical format as shown in \cref{tab:my-table}.
\end{itemize}

In this paper, we present \sysname, \textit{a novel LLM framework specifically designed to mitigate hallucinations and align Verilog code generation with the practices of HDL engineers}. 
The unique challenges in HDL generation, particularly for Verilog, stem from its symbolic nature, knowledge requirements, and the logical complexities inherent in practical design scenarios. 
\sysname addresses these challenges through a three-stage methodology focused on reducing hallucinations in LLM-generated code, while enhancing adherence to the practices of HDL engineers. 
Our key contributions can be summarized as:
\begin{itemize}[leftmargin=*,label={$\triangleright$}] 
    \item \textbf{Mitigating Hallucinations Guided by a Comprehensive Taxonomy:} We introduce a detailed hallucination taxonomy tailored to Verilog code generation (\cref{sec:taxonomy}), encompassing symbolic, knowledge, and logical hallucinations. Using a chain-of-thought (CoT) mechanism (\cref{sec:SI-COT}), \sysname systematically transforms symbolic modalities (\eg, truth tables and state diagrams) into natural language descriptions, guiding the LLM to generate more accurate Verilog code and reducing symbolic hallucination. Besides, we synthesize data that inject domain-specific knowledge (\cref{sec:KE-dataset-gen}) and enhance logical reasoning ability (\cref{sec:LE-dataset-gen}), addressing knowledge and logical hallucinations.
    
    \item \textbf{Aligning Code Generation with the Practices of HDL Engineers:} To the best of our knowledge, \sysname is the first endeavor to bridge the gap between closed-source LLM synthesized data and HDL engineers. We employ a data augmentation strategy that synthesizes high-quality instruction-code pairs by incorporating manual exemplars and synthesized data reflecting key aspects of Verilog design. This enriched dataset ensures that fine-tuned LLMs not only adhere to syntactical correctness, but also align with the intricate requirements of practical HDL engineering scenarios.
    
    \item \textbf{Boosting Verilog Code Generation Performance:} \sysname significantly improves the performance of LLMs in Verilog code generation tasks, achieving up to a 6.7\% increase in pass@1 and a 4.7\% increase in pass@5 compared to competing models on VerilogEval(v1)-Human. By mitigating hallucinations and aligning code generation with the practices of HDL engineers, \sysname enhances the reliability and accuracy of LLM-generated Verilog code, making it more suitable for the use cases of prompts generated by engineers.
    
\end{itemize}


\section{Taxonomy of Hallucination}
\label{sec:taxonomy}
We identify and categorize the hallucinations in threefold, \ie, \emph{symbolic hallucination, knowledge hallucination, and logical hallucination}, as tabulated in \cref{tab:hallucination_taxonomy}.

\paragraph{Symbolic Hallucination} occurs when LLMs encounter symbols, diagrams, or formats that are not directly interpretable. 
LLMs often fail to extract information correctly from these symbolic modalities, causing incorrect or incomplete code generation. 
The symbolic hallucination includes the misinterpretation of state diagrams, waveform charts and truth tables.

\paragraph{Knowledge Hallucination} arises when LLMs lack domain-specific knowledge, leading to irrelevant or incorrect content generation, which can be further specified as:
\begin{itemize}[leftmargin=*,label={$\square$}] 
    \item \underline{Digital Design Convention Misapplication}:
    LLMs are provided with clear module descriptions, but the generated code is poorly structured or incomplete, violating conventions of HDL engineers for commonly used Verilog modules.
    \item \underline{Verilog Syntax Misapplication}:
    LLMs incorrectly use key language constructs, applying them in ways that do not match their intended functionality or meaning.
    \item \underline{Misunderstanding Verilog-specific Attributes}: 
    LLMs struggle to handle the unique characteristics and complex features of Verilog, \eg different reset mechanisms or edge-triggering modes.
\end{itemize}


\paragraph{Logical hallucination} occurs when LLMs lack logical reasoning ability, consequently the output does not follow with the given instructions, which can be further specified as:
\begin{itemize}[leftmargin=*,label={$\square$}] 
    \item \underline{Incorrect Logical Expression}: 
    LLM attempts logical reasoning but produces expressions that do not accurately reflect the requirements of the task, often oversimplifying the logic or failing to satisfy the specific conditions.
    \item \underline{Incorrect handling of corner cases}: 
    LLM generates code that fails to handle corner cases or undefined conditions.
    \item \underline{Failure in adhering to instructional logic}: 
     LLM is expected to faithfully implement the logic described in the instructions but fails, resulting in hallucinations.
\end{itemize}

\section{Methodology}

\begin{figure}[t]
\centering
\includegraphics[width=\linewidth]{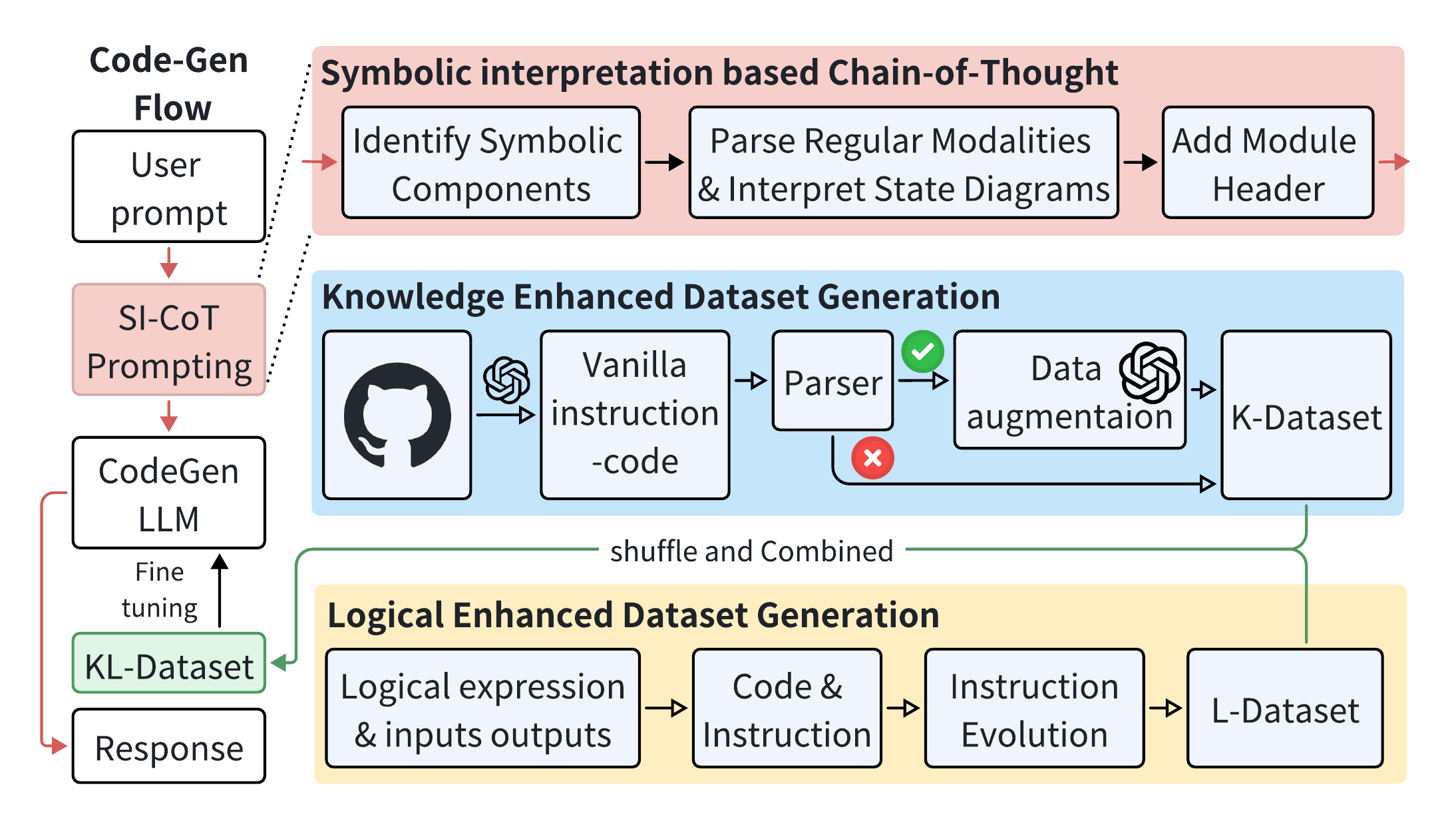}
\caption{\textbf{Overview of \sysname framework}.}
\label{fig:overview}
\end{figure}

\subsection{Overview of \sysname}
The framework overview of \sysname is illustrated in \cref{fig:overview}.
Given a user prompt, \ie, instructions for Verilog code generation (CodeGen), it will be fed into a \emph{CoT prompting model} to translate the vanilla user prompt into a refined CoT prompt, which can mitigate symbolic hallucinations.
Then, the CoT prompt is sent to the \emph{CodeGen-LLM} for an end-to-end inference. In our experiments, we use the same pre-trained models for both \emph{CoT prompting model} and \emph{CodeGen-LLM}.

As countermeasures against the hallucination taxonomy, \sysname integrates three techniques, \ie, \emph{symbolic interpretation based CoT} (SI-CoT) to mitigate symbolic hallucination, while fine-tuning the \emph{CodeGen-LLM} with \emph{knowledge enhanced dataset} (K-dataset) and \emph{logical enhanced dataset} (L-dataset) against knowledge- and logical-hallucinations respectively. 

\subsection{Symbolic Interpretation based Chain-of-Thought}
\label{sec:SI-COT}
SI-CoT leverages structured reasoning to convert symbolic inputs (\ie, state diagrams, waveform charts and truth tables), into an legible form for \emph{CodeGen-LLM}s. 
This approach addresses the challenge of symbolic hallucination, ensuring that the interpretations of input instructions remain aligned with their intended functional descriptions. SI-CoT applies incremental reasoning steps, allowing LLMs to process each input modality systematically, which in \cref{fig:overview} can be further specified as:

\paragraph{\ballnumber{1} Identify Symbolic Components.} 
The first step involves \emph{CoT prompting model} recognizing whether the input contains symbolic representations. 
The process proceeds with interpretation, unless the input consists solely of natural language.

\paragraph{\ballnumber{2} Parse Regular Modalities and Interpret State Diagrams.}
Waveform charts and truth tables are regular modalities that can be directly processed by a parser. 
Once they are identified, an external parser is used to extract relevant data from the representation, then structure them into a uniform instruction format that can be interpreted by the \emph{CodeGen-LLM}. If the identified symbolic component is state diagram, we adopt \emph{CoT prompting model} interprets the diagram by translating its elements into a concise natural language description. 
This description includes the detailed relationships between states, the corresponding outputs, and the transition rules. 
By using a structured prompt template, \emph{CoT prompting model} maintains the integrity of the original instruction while ensuring the symbolic logic is clearly communicated.

\paragraph{\ballnumber{3} Add Module Header.}
Finally, the interpreted instruction is checked to determine whether it contains a complete Verilog module header. If the module header is missing, \emph{CoT prompting model} appends an appropriate module header to define the module name, inputs and outputs, standardizing the instruction.

\cref{tab:cot} shows examples for SI-CoT. Incorporating symbolic interpretation within the CoT framework ensures a more reliable translation of various input modalities, reducing the risk of symbolic hallucination and improving the overall coherence between the input specifications and the generated code.
\begin{table}[t]
\centering
\caption{\textbf{Examples of SI-CoT Interpretation}}
\label{tab:cot}
\resizebox{\columnwidth}{!}{%
\begin{tabular}{ccl}
\hline
\multicolumn{1}{l}{
\begin{tabular}[c]{@{}c@{}}
     \textbf{Instructions before}\\ \textbf{Interpretation} 
\end{tabular}
} & \multicolumn{1}{l}{} & 
\begin{tabular}[c]{@{}c@{}}
     \textbf{Instructions after}\\ 
     \textbf{Interpretation} 
\end{tabular}
\\ \hline
\begin{tabular}[c]{@{}c@{}}A{[}out=0{]}--{[}x=0{]}-\textgreater{}B\\ A{[}out=0{]}--{[}x=1{]}-\textgreater{}A\\ B{[}out=1{]}--{[}x=0{]}-\textgreater{}A\\ B{[}out=1{]}--{[}x=1{]}-\textgreater{}B\end{tabular} & \begin{tabular}[c]{@{}c@{}}
     LLM  \\ $\Rightarrow$
      
\end{tabular}  & \begin{tabular}[c]{@{}l@{}}\underline{\textbf{States\&Outputs}}: 1.state A(out=0); 2. state B(out=1)\\ \underline{\textbf{State transition}}:\\ 1. From state A: If x = 0, then transit to state B; \\ If x = 1, then transit to state A\\ 2. From state B: If x = 0; then transit to state A; \\ If x = 1, then transit to state B\end{tabular} \\ \hline
\begin{tabular}[c]{@{}c@{}}{\;\,  a $\mid$ b $\mid$ out}\\ 0 $\mid$ 0 $\mid$ 0\\ 0 $\mid$ 1 $\mid$ 0\\ 1 $\mid$ 0 $\mid$ 0\\ 1 $\mid$ 1 $\mid$ 1\end{tabular} & \begin{tabular}[c]{@{}c@{}}
     Parser  \\ $\Rightarrow$
      
\end{tabular} 
& \begin{tabular}[c]{@{}l@{}}\underline{\textbf{Variables}}: 1. a(input); 2. b(input); 3. out(output)\\ \underline{\textbf{Rules}}:\\ 1. If a=0, b=0, then out=0;\\ 2. If a=0, b=1, then out=0;\\ 3. If a=1, b=0, then out=0;\\ 4. If a=1, b=1, then out =1;\end{tabular} \\ \hline
\begin{tabular}[c]{@{}c@{}}\qquad a: 0 1 1 0...\\ \qquad b: 1 0 1 0...\\ \qquad out: 1 0 0 1...\\ time(ns): 0 10 20 30 ...\end{tabular} & \begin{tabular}[c]{@{}c@{}}
     Parser  \\ $\Rightarrow$
      
\end{tabular}  & \begin{tabular}[c]{@{}l@{}}\underline{\textbf{Variables}}: 1. a(input); 2. b(input); 3. out(output)\\ \underline{\textbf{Rules}}:\\ When time is 0ns, a=0, b=1, out=1;\\ When time is 10ns, a=1, b=0. out=0...\end{tabular} \\ \bottomrule
\end{tabular}%
}
\end{table}

\begin{figure*}[t]
\centering
\includegraphics[width=0.98\textwidth]{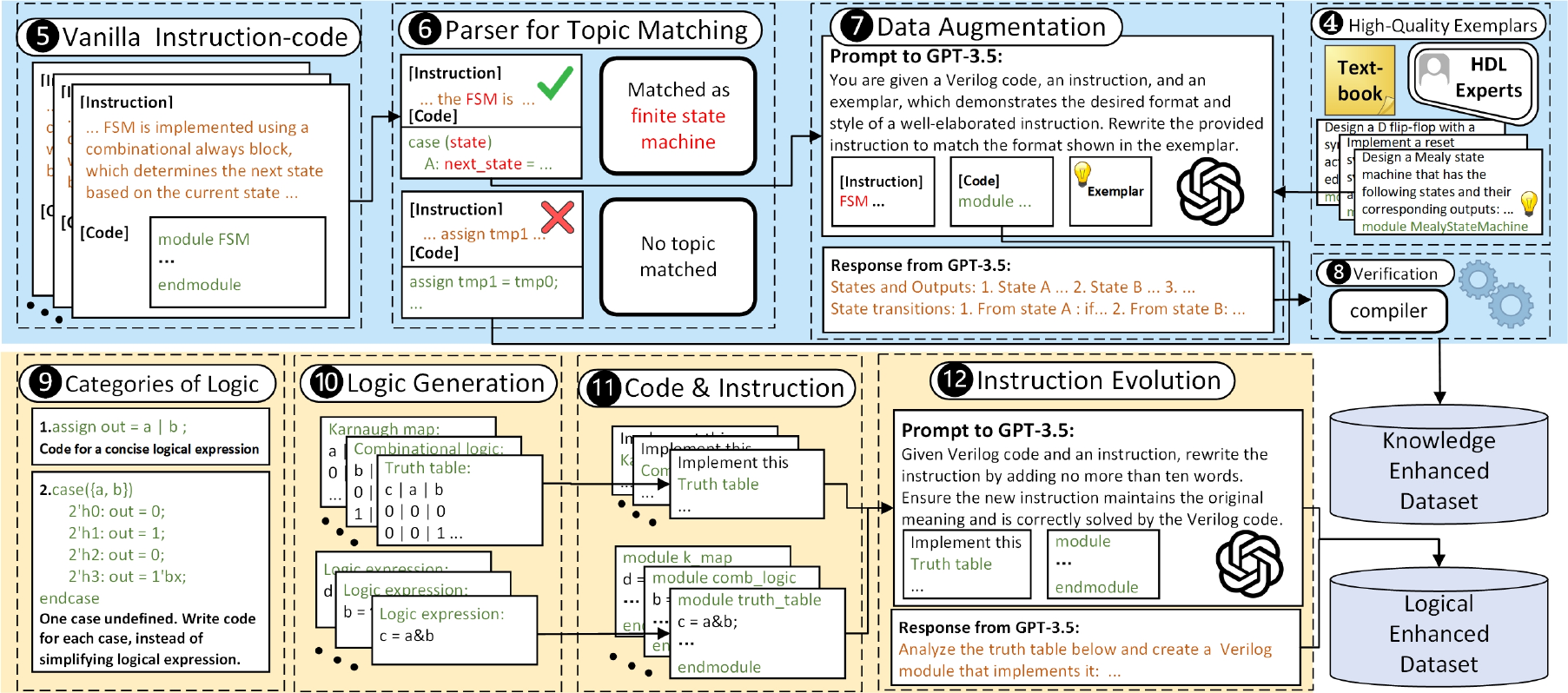}
\caption{\textbf{Generation flow for \emph{knowledge enhanced dataset} (K-dataset) and \emph{logical enhanced dataset} (L-dataset)}. K-dataset and L-dataset are shuffled and combined as KL-dataset to fine-tune a CodeGen-LLM in \sysname.}
\label{fig:dataflow}
\end{figure*}

\subsection{Knowledge-Enhanced Dataset Generation}
\label{sec:KE-dataset-gen}

As shown in the upper (blue shaded) part of \cref{fig:dataflow}, to mitigate knowledge hallucination, we develop a \emph{knowledge enhanced dataset} generation process, which aligns the LLM-generated data with the practices of HDL engineers (human).

\paragraph{\ballnumber{4} High-Quality Exemplars.}  
We curate high-quality exemplars that reflect digital design conventions and Verilog-specific attributes. These exemplars are derived from textbook exercises\cite{textbook1, textbook2,textbook3,Lin2008} and manually designed examples that cover a wide range of Verilog knowledge. 
They include conventions for commonly implemented modules such as finite state machines (FSMs), clock dividers, counters, shift registers, and arithmetic logic units (ALUs). 
Additionally, they incorporate critical Verilog attributes like reset mechanisms (synchronous reset \versus asynchronous reset), clocking and edge sensitivity (positive edge \versus negative edge), and enable signals (active-high \versus active-low enable). 
These exemplars serve as the foundation for constructing HDL-aligned instruction-code pairs, ensuring that the generated instructions follow the questioning style of HDL engineers while embodying high-quality implementations.

\paragraph{\ballnumber{5} Vanilla Instruction-code Pairs.}  
\label{sec:vanilla}
We collect approximately 550,000 Verilog code samples from public GitHub repositories. Following the methodology adopted in previous works\cite{2024origen, autocoder2024}, we use GPT-3.5 to generate vanilla instructions for these code samples. By "vanilla instructions," we refer to basic, general-purpose instructions. These vanilla instruction-code pairs form a foundational dataset, which we refer to as the vanilla dataset. However, they often fall short of the precision and rigor demanded by HDL engineering standards.

\paragraph{\ballnumber{6} Parser for Topic Matching.}  
The parser \texttt{slang}\cite{popoloski2023} is used to identify topics and attributes within the vanilla instruction-code pairs. These topics and attributes are matched with our curated exemplars to ensure each code sample aligns with specific Verilog conventions or attributes. If a vanilla pair lacks identifiable topics or attributes, it can still contribute to the dataset by mitigating the Verilog syntax misapplications.

\paragraph{\ballnumber{7} Data Augmentation.}
Given a related topic exemplar, GPT-3.5 rewrites the vanilla instruction to closely align it with the high-quality exemplar. If a vanilla instruction is associated with multiple exemplars, it is rewritten separately for each exemplar.

\paragraph{\ballnumber{8} Verification.}  
To ensure the validity of the generated instruction-code pairs, we compile the Verilog code using an industry-standard Verilog compiler.
This verification step ensures erroneous or incomplete pairs are filtered out.

\subsection{Logical Enhanced Dataset Generation}
\label{sec:LE-dataset-gen}

As shown in the lower (yellow shaded) part of \cref{fig:dataflow}, we develop an L-dataset generation process to mitigate logical hallucination and enhance accurate logical reasoning. 

\paragraph{\ballnumber{9} Two Categories of Logical Reasoning in Verilog.}
In Verilog, logical reasoning comes in twofold: finding the most concise logical expression, and faithfully implementing the logic when no concise expression is available. We generate datasets that cover both scenarios, allowing the fine-tuned model to select the appropriate approach during inference.

\paragraph{\ballnumber{10} Generate Logical Expressions and Input-Output Values.} 
We begin by developing scripts that produce a wide range of logical expressions and their associated input-output mappings. The generated input-output values are assigned with a typical logic problem encountered in Verilog, \eg, Karnaugh maps and combinational logic. These values serve as a basis for ensuring that the model understands the logical relationships in Verilog.

\paragraph{\ballnumber{11} Integration of code and instructions.} 
The next step is to incorporate the generated logical expressions and input-output values into pre-designed code templates and instruction templates. By embedding logical expressions into these templates, we create scenarios that reflect both straightforward and complex logical reasoning tasks.

\begin{table*}[t]
\centering
\caption{\textbf{Comparison of \sysname against various baseline models}. }
\label{tab:comparison}
\begin{tabular}{crcccccccccc}
\cline{2-12}
\multicolumn{1}{l|}{} &
  \multicolumn{1}{c|}{} &
  \multicolumn{1}{c|}{} &
  \multicolumn{1}{c|}{} &
  \multicolumn{4}{c|}{VerilogEval v1 (\%)} &
  \multicolumn{2}{c|}{RTLLM v1.1 (\%)} &
  \multicolumn{2}{c|}{VerilogEval v2 (\%)} \\ \cline{5-12} 
\multicolumn{1}{l|}{} &
  \multicolumn{1}{c|}{} &
  \multicolumn{1}{c|}{} &
  \multicolumn{1}{c|}{} &
  \multicolumn{2}{c|}{} &
  \multicolumn{2}{c|}{} &
  \multicolumn{1}{c|}{} &
  \multicolumn{1}{c|}{} &
  \multicolumn{2}{c|}{Task: Spec.-to-RTL} \\ \cline{11-12} 
\multicolumn{1}{l|}{} &
  \multicolumn{1}{c|}{} &
  \multicolumn{1}{c|}{} &
  \multicolumn{1}{c|}{} &
  \multicolumn{2}{c|}{\multirow{-2}{*}{Machine}} &
  \multicolumn{2}{c|}{\multirow{-2}{*}{Human}} &
  \multicolumn{1}{c|}{} &
  \multicolumn{1}{c|}{} &
  \multicolumn{2}{c|}{0-Shot (n=1)} \\ \cline{5-8} \cline{11-12} 
\multicolumn{1}{c|}{} &
  \multicolumn{1}{c|}{\multirow{-4}{*}{Model}} &
  \multicolumn{1}{c|}{\multirow{-4}{*}{\begin{tabular}[c]{@{}c@{}}Open\\ source\end{tabular}}} &
  \multicolumn{1}{c|}{\multirow{-4}{*}{\begin{tabular}[c]{@{}c@{}}Model\\ Size\end{tabular}}} &
  \multicolumn{1}{c|}{pass@1} &
  \multicolumn{1}{c|}{pass@5} &
  \multicolumn{1}{c|}{pass@1} &
  \multicolumn{1}{c|}{pass@5} &
  \multicolumn{1}{c|}{\multirow{-3}{*}{\begin{tabular}[c]{@{}c@{}}Syntax\\ pass@5\end{tabular}}} &
  \multicolumn{1}{c|}{\multirow{-3}{*}{\begin{tabular}[c]{@{}c@{}}Func.\\ pass@5\end{tabular}}} &
  \multicolumn{1}{c|}{pass@1} &
  \multicolumn{1}{c|}{pass@5} \\ \hline
\multicolumn{1}{|c|}{} &
  \multicolumn{1}{r|}{GPT-3.5} &
  \multicolumn{1}{c|}{\xmark} &
  \multicolumn{1}{c|}{n/a} &
  \multicolumn{1}{c|}{46.7} &
  \multicolumn{1}{c|}{69.1} &
  \multicolumn{1}{c|}{26.7} &
  \multicolumn{1}{c|}{45.8} &
  \multicolumn{1}{c|}{89.7} &
  \multicolumn{1}{c|}{37.9} &
  \multicolumn{1}{c|}{n/a} &
  \multicolumn{1}{c|}{n/a} \\
\multicolumn{1}{|c|}{} &
  \multicolumn{1}{r|}{GPT-4} &
  \multicolumn{1}{c|}{\xmark} &
  \multicolumn{1}{c|}{n/a} &
  \multicolumn{1}{c|}{60.0} &
  \multicolumn{1}{c|}{70.6} &
  \multicolumn{1}{c|}{43.5} &
  \multicolumn{1}{c|}{55.8} &
  \multicolumn{1}{c|}{\cellcolor[HTML]{FFD800}100.0} &
  \multicolumn{1}{c|}{\cellcolor[HTML]{C0C0C0}65.5} &
  \multicolumn{1}{c|}{44.2} &
  \multicolumn{1}{c|}{n/a} \\
\multicolumn{1}{|c|}{} &
  \multicolumn{1}{r|}{Starcoder\cite{li2023starcoder}} &
  \multicolumn{1}{c|}{\cmark} &
  \multicolumn{1}{c|}{15B} &
  \multicolumn{1}{c|}{46.8} &
  \multicolumn{1}{c|}{54.5} &
  \multicolumn{1}{c|}{18.1} &
  \multicolumn{1}{c|}{26.1} &
  \multicolumn{1}{c|}{93.1} &
  \multicolumn{1}{c|}{27.6} &
  \multicolumn{1}{c|}{n/a} &
  \multicolumn{1}{c|}{n/a} \\
\multicolumn{1}{|c|}{} &
  \multicolumn{1}{r|}{CodeLlama\cite{rozière2024code}} &
  \multicolumn{1}{c|}{\cmark} &
  \multicolumn{1}{c|}{7B} &
  \multicolumn{1}{c|}{43.1} &
  \multicolumn{1}{c|}{47.1} &
  \multicolumn{1}{c|}{18.2} &
  \multicolumn{1}{c|}{22.7} &
  \multicolumn{1}{c|}{86.2} &
  \multicolumn{1}{c|}{31.0} &
  \multicolumn{1}{c|}{n/a} &
  \multicolumn{1}{c|}{n/a} \\
\multicolumn{1}{|c|}{} &
  \multicolumn{1}{r|}{DeepSeek-Coder\cite{guo2024deepseekcoderlargelanguagemodel}} &
  \multicolumn{1}{c|}{\cmark} &
  \multicolumn{1}{c|}{6.7B} &
  \multicolumn{1}{c|}{52.2} &
  \multicolumn{1}{c|}{55.4} &
  \multicolumn{1}{c|}{30.2} &
  \multicolumn{1}{c|}{33.9} &
  \multicolumn{1}{c|}{93.1} &
  \multicolumn{1}{c|}{44.8} &
  \multicolumn{1}{c|}{28.2} &
  \multicolumn{1}{c|}{n/a} \\
\multicolumn{1}{|c|}{\multirow{-6}{*}{\rot{General LLM}}} &
  \multicolumn{1}{r|}{CodeQwen\cite{bai2023qwentechnicalreport}} &
  \multicolumn{1}{c|}{\cmark} &
  \multicolumn{1}{c|}{7B} &
  \multicolumn{1}{c|}{46.5} &
  \multicolumn{1}{c|}{54.9} &
  \multicolumn{1}{c|}{22.5} &
  \multicolumn{1}{c|}{26.1} &
  \multicolumn{1}{c|}{86.2} &
  \multicolumn{1}{c|}{41.4} &
  \multicolumn{1}{c|}{n/a} &
  \multicolumn{1}{c|}{n/a} \\ \hline
\multicolumn{1}{|c|}{} &
  \multicolumn{1}{r|}{ChipNeMo\cite{liu2024chipnemo}} &
  \multicolumn{1}{c|}{\xmark} &
  \multicolumn{1}{c|}{13B} &
  \multicolumn{1}{c|}{43.4} &
  \multicolumn{1}{c|}{n/a} &
  \multicolumn{1}{c|}{22.4} &
  \multicolumn{1}{c|}{n/a} &
  \multicolumn{1}{c|}{n/a} &
  \multicolumn{1}{c|}{n/a} &
  \multicolumn{1}{c|}{n/a} &
  \multicolumn{1}{c|}{n/a} \\
\multicolumn{1}{|c|}{} &
  \multicolumn{1}{r|}{Thakur et al.\cite{thakur2024verigen}} &
  \multicolumn{1}{c|}{\cmark} &
  \multicolumn{1}{c|}{16B} &
  \multicolumn{1}{c|}{44.0} &
  \multicolumn{1}{c|}{52.6} &
  \multicolumn{1}{c|}{30.3} &
  \multicolumn{1}{c|}{43.9} &
  \multicolumn{1}{c|}{86.2} &
  \multicolumn{1}{c|}{24.1} &
  \multicolumn{1}{c|}{n/a} &
  \multicolumn{1}{c|}{n/a} \\
\multicolumn{1}{|c|}{} &
  \multicolumn{1}{r|}{RTLCoder-Mistral\cite{liu2024rtlcoder}} &
  \multicolumn{1}{c|}{\cmark} &
  \multicolumn{1}{c|}{7B} &
  \multicolumn{1}{c|}{62.5} &
  \multicolumn{1}{c|}{72.2} &
  \multicolumn{1}{c|}{36.7} &
  \multicolumn{1}{c|}{45.5} &
  \multicolumn{1}{c|}{\cellcolor[HTML]{C0C0C0}96.6} &
  \multicolumn{1}{c|}{48.3} &
  \multicolumn{1}{c|}{n/a} &
  \multicolumn{1}{c|}{n/a} \\
\multicolumn{1}{|c|}{} &
  \multicolumn{1}{r|}{RTLCoder-DeepSeek\cite{liu2024rtlcoder}} &
  \multicolumn{1}{c|}{\cmark} &
  \multicolumn{1}{c|}{6.7B} &
  \multicolumn{1}{c|}{61.2} &
  \multicolumn{1}{c|}{76.5} &
  \multicolumn{1}{c|}{41.6} &
  \multicolumn{1}{c|}{50.1} &
  \multicolumn{1}{c|}{93.1} &
  \multicolumn{1}{c|}{48.3} &
  \multicolumn{1}{c|}{36.5} &
  \multicolumn{1}{c|}{n/a} \\
\multicolumn{1}{|c|}{} &
  \multicolumn{1}{r|}{BetterV-CodeLlama\cite{pei2024betterv}} &
  \multicolumn{1}{c|}{\xmark} &
  \multicolumn{1}{c|}{7B} &
  \multicolumn{1}{c|}{64.2} &
  \multicolumn{1}{c|}{75.4} &
  \multicolumn{1}{c|}{40.9} &
  \multicolumn{1}{c|}{50.0} &
  \multicolumn{1}{c|}{n/a} &
  \multicolumn{1}{c|}{n/a} &
  \multicolumn{1}{c|}{n/a} &
  \multicolumn{1}{c|}{n/a} \\
\multicolumn{1}{|c|}{} &
  \multicolumn{1}{r|}{BetterV-DeepSeek\cite{pei2024betterv}} &
  \multicolumn{1}{c|}{\xmark} &
  \multicolumn{1}{c|}{6.7B} &
  \multicolumn{1}{c|}{67.8} &
  \multicolumn{1}{c|}{79.1} &
  \multicolumn{1}{c|}{45.9} &
  \multicolumn{1}{c|}{53.3} &
  \multicolumn{1}{c|}{n/a} &
  \multicolumn{1}{c|}{n/a} &
  \multicolumn{1}{c|}{n/a} &
  \multicolumn{1}{c|}{n/a} \\
\multicolumn{1}{|c|}{} &
  \multicolumn{1}{r|}{BetterV-CodeQwen\cite{pei2024betterv}} &
  \multicolumn{1}{c|}{\xmark} &
  \multicolumn{1}{c|}{7B} &
  \multicolumn{1}{c|}{68.1} &
  \multicolumn{1}{c|}{79.4} &
  \multicolumn{1}{c|}{46.1} &
  \multicolumn{1}{c|}{53.7} &
  \multicolumn{1}{c|}{n/a} &
  \multicolumn{1}{c|}{n/a} &
  \multicolumn{1}{c|}{n/a} &
  \multicolumn{1}{c|}{n/a} \\
\multicolumn{1}{|c|}{} &
  \multicolumn{1}{r|}{AutoVCoder-CodeLlama\cite{autocoder2024}} &
  \multicolumn{1}{c|}{\xmark} &
  \multicolumn{1}{c|}{7B} &
  \multicolumn{1}{c|}{63.7} &
  \multicolumn{1}{c|}{72.9} &
  \multicolumn{1}{c|}{44.5} &
  \multicolumn{1}{c|}{52.8} &
  \multicolumn{1}{c|}{93.1} &
  \multicolumn{1}{c|}{48.3} &
  \multicolumn{1}{c|}{n/a} &
  \multicolumn{1}{c|}{n/a} \\
\multicolumn{1}{|c|}{} &
  \multicolumn{1}{r|}{AutoVCoder-DeepSeek\cite{autocoder2024}} &
  \multicolumn{1}{c|}{\xmark} &
  \multicolumn{1}{c|}{6.7B} &
  \multicolumn{1}{c|}{69.0} &
  \multicolumn{1}{c|}{79.3} &
  \multicolumn{1}{c|}{46.9} &
  \multicolumn{1}{c|}{53.7} &
  \multicolumn{1}{c|}{\cellcolor[HTML]{FFD800}100.0} &
  \multicolumn{1}{c|}{51.7} &
  \multicolumn{1}{c|}{n/a} &
  \multicolumn{1}{c|}{n/a} \\
\multicolumn{1}{|c|}{} &
  \multicolumn{1}{r|}{AutoVCoder-CodeQwen\cite{autocoder2024}} &
  \multicolumn{1}{c|}{\xmark} &
  \multicolumn{1}{c|}{7B} &
  \multicolumn{1}{c|}{68.7} &
  \multicolumn{1}{c|}{79.9} &
  \multicolumn{1}{c|}{48.5} &
  \multicolumn{1}{c|}{55.9} &
  \multicolumn{1}{c|}{\cellcolor[HTML]{FFD800}100.0} &
  \multicolumn{1}{c|}{51.7} &
  \multicolumn{1}{c|}{n/a} &
  \multicolumn{1}{c|}{n/a} \\
\multicolumn{1}{|c|}{\multirow{-11}{*}{\rot{LLM for Verilog CodeGen}}} &
  \multicolumn{1}{r|}{OriGen-DeepSeek-7B-v1.5\textsuperscript{$\star$}\cite{2024origen}} &
  \multicolumn{1}{c|}{\cmark} &
  \multicolumn{1}{c|}{7B} &
  \multicolumn{1}{c|}{74.1} &
  \multicolumn{1}{c|}{\cellcolor[HTML]{C0C0C0}82.4} &
  \multicolumn{1}{c|}{\cellcolor[HTML]{CD7F32}54.4} &
  \multicolumn{1}{c|}{\cellcolor[HTML]{CD7F32}60.1} &
  \multicolumn{1}{c|}{n/a} &
  \multicolumn{1}{c|}{\cellcolor[HTML]{C0C0C0}65.5} &
  \multicolumn{1}{c|}{n/a} &
  \multicolumn{1}{c|}{n/a} \\ \hline
\multicolumn{1}{|c|}{} &
  \multicolumn{1}{r|}{\sysname-CodeLlama} &
  \multicolumn{1}{c|}{\cmark} &
  \multicolumn{1}{c|}{7B} &
  \multicolumn{1}{c|}{\cellcolor[HTML]{CD7F32}74.7} &
  \multicolumn{1}{c|}{80.0} &
  \multicolumn{1}{c|}{51.3} &
  \multicolumn{1}{c|}{59.0} &
  \multicolumn{1}{c|}{\cellcolor[HTML]{CD7F32}95.4} &
  \multicolumn{1}{c|}{54.7} &
  \multicolumn{1}{c|}{\cellcolor[HTML]{CD7F32}46.4} &
  \multicolumn{1}{c|}{\cellcolor[HTML]{CD7F32}55.8} \\
\multicolumn{1}{|c|}{} &
  \multicolumn{1}{r|}{\sysname-DeepSeek} &
  \multicolumn{1}{c|}{\cmark} &
  \multicolumn{1}{c|}{6.7B} &
  \multicolumn{1}{c|}{\cellcolor[HTML]{FFD800}78.8} &
  \multicolumn{1}{c|}{\cellcolor[HTML]{FFD800}84.5} &
  \multicolumn{1}{c|}{\cellcolor[HTML]{C0C0C0}57.3} &
  \multicolumn{1}{c|}{\cellcolor[HTML]{C0C0C0}64.2} &
  \multicolumn{1}{c|}{92.8} &
  \multicolumn{1}{c|}{\cellcolor[HTML]{FFD800}66.0} &
  \multicolumn{1}{c|}{\cellcolor[HTML]{FFD800}58.3} &
  \multicolumn{1}{c|}{\cellcolor[HTML]{FFD800}63.4} \\
\multicolumn{1}{|c|}{\multirow{-3}{*}{\rot{\textbf{Ours}}}} &
  \multicolumn{1}{r|}{\sysname-CodeQwen} &
  \multicolumn{1}{c|}{\cmark} &
  \multicolumn{1}{c|}{7B} &
  \multicolumn{1}{c|}{\cellcolor[HTML]{C0C0C0}77.3} &
  \multicolumn{1}{c|}{\cellcolor[HTML]{CD7F32}81.2} &
  \multicolumn{1}{c|}{\cellcolor[HTML]{FFD800}61.1} &
  \multicolumn{1}{c|}{\cellcolor[HTML]{FFD800}64.8} &
  \multicolumn{1}{c|}{92.8} &
  \multicolumn{1}{c|}{\cellcolor[HTML]{CD7F32}62.2} &
  \multicolumn{1}{c|}{\cellcolor[HTML]{C0C0C0}54.6} &
  \multicolumn{1}{c|}{\cellcolor[HTML]{C0C0C0}62.9} \\ \hline
\multicolumn{12}{l}{
$\dagger$: ranking of performance as \textcolor{gold}{\faTrophy} 1st, \textcolor{silver}{\faTrophy} 2nd, \textcolor{bronze}{\faTrophy} 3rd place; \hspace{1em}
} \\
\multicolumn{12}{l}{
$\star$: Only the base model of OriGen is DeepSeek-v1.5(7B), while the other works utilize DeepSeek-v1(6.7B)
} \\
\multicolumn{12}{l}{
Note: To the best of our knowledge, ChipNeMo \cite{liu2024chipnemo}, BetterV \cite{pei2024betterv}, and AutoVCoder \cite{autocoder2024} are not publicly accessible at the time of writing.
}
\end{tabular}
\end{table*}
\paragraph{\ballnumber{12} Instruction Evolution.}
Instruction evolution\cite{wizardlm2024} is frequently employed to generate diverse and complex instructions. To introduce linguistic variety, we use GPT-3.5 to rewrite the original instructions while ensuring the semantic core is retained. The modifications are constrained to adding or removing no more than ten words, preserving the logical structure. This step enhances the diversity of the dataset, improving the robustness of the model against overfitting.

By following the generation flow shown in \cref{fig:dataflow}, we obtain approximately 43k valid vanilla instruction-code pairs, 14k HDL-aligned instruction-code pairs for mitigating knowledge hallucination (\ie, K-dataset), and 5k logically enhanced instruction-code pairs for mitigating logical hallucination (\ie, L-dataset). They are used to fine-tune the \emph{CodeGen-LLM}.

\section{Experiments}


\subsection{Experiment setup}
\paragraph{Models and Platform.}
We utilize CodeLlama-7b-Instruct\cite{rozière2024code}, Deepseek-Coder-6.7b-Instruct\cite{guo2024deepseekcoderlargelanguagemodel}, and CodeQwen1.5-7B-Chat\cite{bai2023qwentechnicalreport} as base models. One model is used for SI-CoT , fine-tuning and code generation.
Fine-tuning is performed on two Nvidia A100-80GB GPUs for 3 epochs. We apply the AdamW optimizer with a cosine learning rate scheduler, initializing with 15 warm-up iterations with a learning rate of 5e−5. 
The global batch size is set to 256.
Following RTLCoder \cite{liu2024rtlcoder}, we set the temperature of each model to 0.2, 0.5 and 0.8, reporting the best performance. 

\paragraph{Benchmark and Evaluation Metrics.}
Evaluations are performed on: VerilogEval v1\cite{liu2023verilogeval}, RTLLM v1.1 \cite{lu2024rtllm}, and the newly introduced VerilogEval v2\cite{pinckney2024revisiting}. 
VerilogEval v1 is divided into two distinct components:
VerilogEval-machine, which includes 143 Verilog generation tasks created using GPT-based models, and VerilogEval-human, a collection of 156 manually crafted problems.
RTLLM v1.1 consists of 29 tasks focused on RTL design.
VerilogEval v2 builds on the original VerilogEval-Human, extending its scope to encompass specification-to-RTL design tasks\cite{pinckney2024revisiting}, whose  prompt style is as a chat bot, with well-defined ``Question” and ``Answer”.

We evaluate design performance from two key perspectives: \emph{syntax correctness} and \emph{functional correctness}. Both correctness are measured by the pass@k metric\cite{chen2021evaluatinglargelanguagemodels}, which estimates the proportion of problems that can be solved in at least one of k attempts ($k \in \{1,5
\}$). It is expressed as:
\begin{equation}
\text{pass}@k := \E\left[1 - \binom{n - c}{k}/\binom{n}{k}\right]    
\end{equation}
where \(n \geq k\) denotes the total number of trials per problem, and \(c\) represents the number of trials that passed the functional check. We set \(n = 10\) in experiments.

\subsection{Comparison with Competing Works}
We evaluate the syntactic and functional correctness of \sysname, which is open-source and no larger than 7B, on three benchmarks. Results are tabulated in \cref{tab:comparison}, which indicate \sysname performs best on three benchmarks in functional correctness. 
Specifically, \sysname-DeepSeek performs 4.7\% and 2.1\% higher than OriGen in pass@1 and pass@5 respectively on VerilogEval-Machine.
\sysname-CodeQwen 
outperforms OriGen by 6.7\% and 4.7\% in pass@1 and pass@5 respectively on VerilogEval-Human, whose tasks are more aligned with HDL engineers and more closer to real-world practices. 
\sysname-DeepSeek on RTLLM v1.1 performs the best 
in
functional pass@5, which outperforms both OriGen and GPT-4 by 0.5\%. Although AutoVCoder\cite{autocoder2024} is fine-tuned on a larger dataset based on around one million Verilog modules and 50,000 synthetic code samples, our \sysname-DeepSeek, fine-tuned on just 62,000 data samples, achieves only 4-6\% lower in terms of syntax pass@5.
Our models also perform well on VerilogEval v2, whose prompts are more conform to expectations of HDL engineers. \sysname-DeepSeek achieves 58.3\% and 63.4\% in pass@1 and pass@5 respectively. Additionally, after fine-tuning, CodeLlama performs worse than the other two models, which aligns with the experimental results reported in other study\cite{autocoder2024}. These results demonstrate that \sysname exhibits significant superiority in Verilog code generation and meets practical needs of HDL engineers.

\subsection{Evaluation on Symbolic Modalities}
\label{subsec:evaluation_non_natural_language}
To evaluate how \sysname handles symbolic modalities, we curate a set of 44 tasks from VerilogEval(v1)-Human, which span three categories that frequently induce hallucinations: \textit{truth tables(10)}, \textit{waveform charts(13)}, and \textit{state diagrams(21)}.

In this experiment, we compare our model, HaVen-CodeQwen, against commercial LLMs (GPT-4 and DeepSeek-Coder-V2\cite{deepseekai2024deepseekcoderv2breakingbarrierclosedsource}) and open-source Verilog CodeGen models (RTLCoder\cite{liu2024rtlcoder} and OriGen\cite{2024origen}), on the set of 44 tasks.

The results, shown in \cref{tab:modality_results_transposed}, reveal an interesting trend: GPT-4 and OriGen exhibit similar performance, while DeepSeek-Coder-V2 outperforms both in all three categories, though it remains second to our model. By breaking down performance in each modality, \sysname shows its superior performance in handling these symbolic modalities, a common pain point for LLMs. 
\sysname-CodeQwen outperforms all models in three categories, achieving the highest overall pass@1 of 47.7\%. This underscores the effectiveness of \sysname in generating accurate Verilog code from complex instructions.

\begin{table}[]
\centering
\caption{\textbf{Evaluation on Symbolic Modalities}. P/T denotes pass cases/total cases. PR is the pass rate.}
\label{tab:modality_results_transposed}
\resizebox{\columnwidth}{!}{%
\begin{tabular}{|l|c|c|c|c|}
\hline
\multicolumn{1}{|c|}{\multirow{2}{*}{Model}} & Truth Table & Waveform & State Diagram & \multirow{2}{*}{\begin{tabular}[c]{@{}c@{}}Overall\\ Pass@1\end{tabular}} \\ \cline{2-4}
\multicolumn{1}{|c|}{}         & P/T (PR) & P/T (PR) & P/T (PR) &  \\ \hline
RTLCoder\cite{liu2024rtlcoder} &    1/10(10.0\%)  &  3/13(23.1\%)   &  3/21 (14.3\%)       & 15.9\% \\ \hline
OriGen\cite{2024origen}        & 2/10(20.0\%)   & 3/13(23.1\%)  & 5/21(23.8\%)   & 22.7\% \\ \hline
GPT-4     &  2/10(20.0\%)    & 3/13(23.1\%)  & 5/21(23.8\%)      & 22.7\% \\ \hline
DeepSeek-Coder-V2    & 3/10(30.0\%)    & 3/13(23.1\%)         &   9/21(42.9\%)       & 34.1\% \\ \hline
\sysname-CodeQwen  & 6/10(60.0\%)    & 4/13(30.8\%)         &   11/21(52.4\%)       & 47.4\% \\ \hline
\end{tabular}%
}
\end{table}


\subsection{Ablation study}

\begin{figure}[t]
\centering
\includegraphics[width=\linewidth]{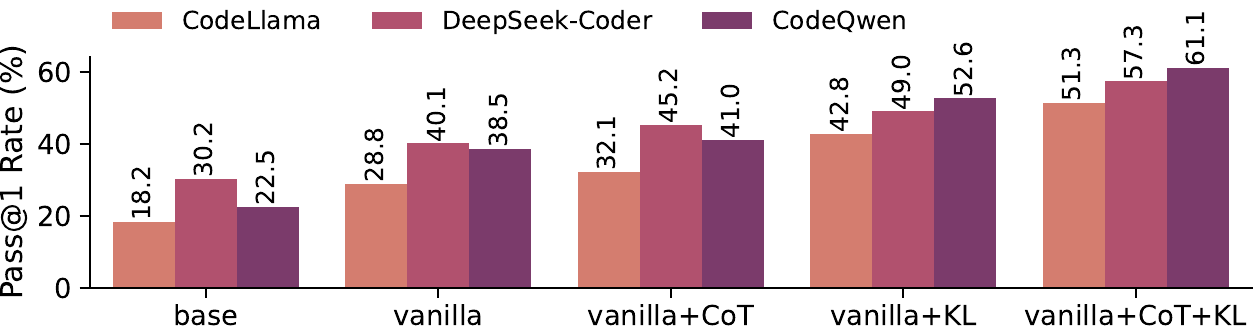} \\ \vspace{.5em}
\includegraphics[width=\linewidth]{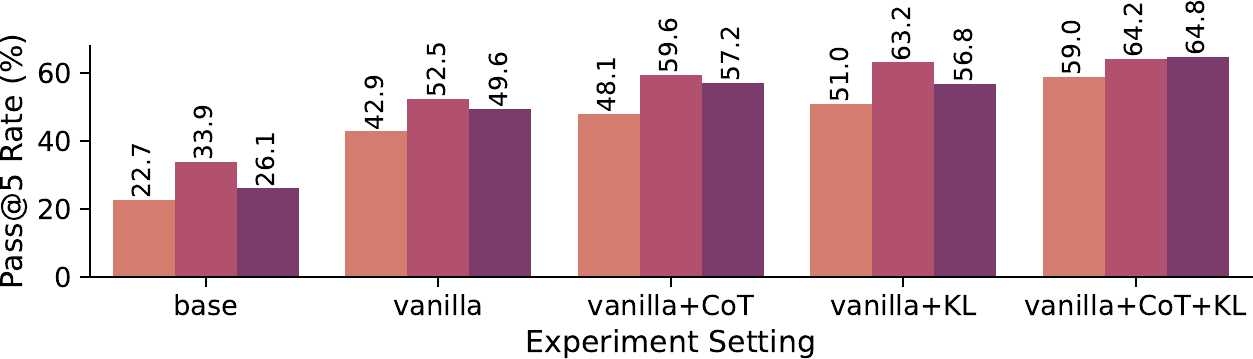}
\caption{\textbf{Ablation Study of techniques adopted in \sysname}. Pass@1/5 are reported on VerilogEval(v1)-Human dataset.}
\label{fig:ablation_of_techniques}
\end{figure}

\paragraph{Effectiveness of Techniques} is evaluated on the VerilogEval(v1)-Human (\cref{fig:ablation_of_techniques}), with five settings:
\begin{itemize}[leftmargin=*,label={$\square$}] 
    \item \underline{Base}: Original pre-trained LLMs with no modifications;
    \item \underline{Vanilla}: Base LLMs fine-tuned only by vanilla dataset (\cref{sec:vanilla}).
    \item \underline{Vanilla+CoT}: Base LLMs fine-tuned by vanilla dataset, then prompted by SI-CoT instructions.
    \item \underline{Vanilla+KL}: CodeGen LLMs fine-tuned by KL-dataset.
    \item \underline{Vanilla+CoT+KL}: CodeGen LLMs fine-tuned by KL-dataset, then prompted by SI-CoT instructions.
\end{itemize}
The incorporation of the proposed SI-CoT and KL-dataset in \sysname shows effectiveness across three base LLMs.
Solely adopting SI-CoT improve the Pass@1 and Pass@5 by 3.6\% and 6.6\% on average, while fine-tuning LLMs with KL-dataset improves these metrics by 12.3 \% and 8.7\% on average. When combining SI-CoT and KL-dataset together, we still observe an improvement, which demonstrates that they supplement each other.


\begin{table}[t]
\centering
\caption{\textbf{Evaluation of SI-CoT on commercial LLMs.}}
\label{tab:cot_on_commercialLLM}
\resizebox{\columnwidth}{!}{%
\begin{tabular}{c|c|c|c|}
\cline{2-4}
                                          & GPT-4o mini & GPT-4 & DeepSeek-Coder-V2 \\ \hline
\multicolumn{1}{|c|}{Pass@1 (w SI-CoT)}   & 22.7\%       & 22.7\% & 34.1\%             \\ \hline
\multicolumn{1}{|c|}{Pass@1 (w/o SI-CoT)} & 31.8\%       & 34.1\% & 45.5\%             \\ \hline
\end{tabular}%
}
\end{table}

\paragraph{Analysis of SI-CoT}
is further extended to commercial LLMs to assess the effectiveness of SI-CoT.
We select a set of 44 tasks from
VerilogEval(v1)-Human, same as we conducted in \cref{tab:modality_results_transposed}. 
For fair comparison, all LLMs are prompted with the same SI-CoT instructions produced by the base model CodeQwen\cite{bai2023qwentechnicalreport}. 
The results in \cref{tab:cot_on_commercialLLM} shows SI-CoT directly helps with CodeGen LLM even without fine-tuning.
Moreover, the results reveal an intriguing finding: GPT-4o mini, despite being cheaper than GPT-4, delivers comparable performance. 
Meanwhile, DeepSeek-Coder-V2, the most affordable of the three commercial LLMs, demonstrates impressive results, matching GPT-4's performance with SI-CoT, even without using SI-CoT.


\begin{figure}[t]
\centering
\includegraphics[width=\linewidth]{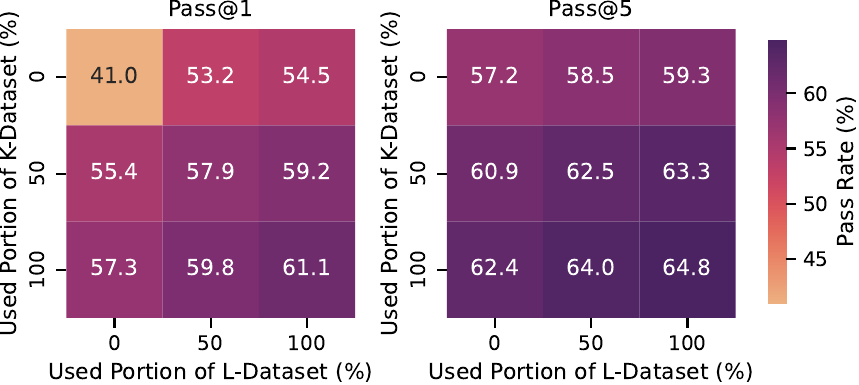}
\caption{\textbf{Ablation Study of composition of KL-dataset}.}
\label{fig:ablation_of_datasets}
\end{figure}

\paragraph{Analysis of KL-dataset.}
As depicted in \cref{fig:ablation_of_datasets}, we use \{0\%, 50\%, 100\%\} portions from K-dataset and L-dataset and mix them to fine-tune the \emph{CodeGen-LLM} (\ie, CodeQwen).
Such experiment is to evaluate the effectiveness of K-dataset and L-dataset, using VerilogEval(v1)-Human.
The results reveal that both K-dataset and L-dataset contribute to the pass@k improvement.
The improvement resulting from the K-dataset is higher, where the K-dataset contains more data than L-dataset could be the reason.
In addition, further enlarging the samples in KL-dataset can still be beneficial to optimize \sysname.


\section{Conclusion}
In this work, we present \sysname, a novel LLM framework for Verilog code generation. We introduce a hallucination taxonomy, including symbolic, knowledge and logical hallucinations. \sysname addresses these challenges through a three-stage methodology focused on reducing hallucinations in LLM-generated code while enhancing adherence to the practices of HDL engineers. Experimental results demonstrate that \sysname outperforms existing state-of-the-art methods and that our three techniques are effective individually. 


\newpage
\bibliographystyle{IEEEtran}
\bibliography{reference}

\end{document}

%% file: math_commands.tex
\usepackage{amsmath,amsfonts,bm}

\def\eqref#1{equation~\ref{#1}}

\def\1{\bm{1}}

\DeclareMathAlphabet{\mathsfit}{\encodingdefault}{\sfdefault}{m}{sl}
\SetMathAlphabet{\mathsfit}{bold}{\encodingdefault}{\sfdefault}{bx}{n}

\newcommand{\E}{\mathbb{E}}

